\input{aipcheck}

\documentclass[
  ,final            % use final for the camera ready runs
  ]
  {aipproc}

\layoutstyle{8x11single}

\usepackage{amssymb}
\usepackage{graphicx}
\usepackage{amsmath}%
\usepackage{amsfonts}

\begin{document}

\title{Phase Transitions and Quantum Measurements}


\author{Armen E. Allahverdyan}{
  address={Yerevan Physics Institute, Alikhanian Brothers St. 2, Yerevan 375036, Armenia}
}

\author{Roger Balian}{
  address={Service de Physique Th\'{e}orique, CEA/Saclay - DSM/SPhT, F-91191 Gif sur Yvette Cedex, France}
  \thanks{
Invited speaker at the conference QTRF3, V\"{a}xj\"{o}, Sweden, 6-11 june 2005. E-mail: balian@cea.fr}
}

\author{Theo M. Nieuwenhuizen}{
  address={Institute for Theoretical Physics, Valckenierstraat 65, 1018 XE Amsterdam, The Netherlands}
}



\begin{abstract}
In a quantum measurement, a coupling $g$ between the system $\mathrm{S}$ and
the apparatus $\mathrm{A}$ triggers the establishment of correlations, which
provide statistical information about $\mathrm{S}$. Robust registration
requires $\mathrm{A}$ to be macroscopic, and a dynamical symmetry breaking of
$\mathrm{A}$ governed by $\mathrm{S}$ allows the absence of any bias. Phase
transitions are thus a paradigm for quantum measurement apparatuses, with the
order parameter as pointer variable. The coupling $g$ behaves as the source of
symmetry breaking. The exact solution of a model where $\mathrm{S}$ is a
single spin and $\mathrm{A}$ a magnetic dot (consisting of $N$ interacting spins
and a phonon thermal bath) exhibits the reduction of the state as a relaxation
process of the off-diagonal elements of $\mathrm{S}+\mathrm{A}$, rapid due to
the large size of $N$. The registration of the diagonal elements involves a
slower relaxation from the initial paramagnetic state of $\mathrm{A}$ to
either one of its ferromagnetic states. If $g$ is too weak, the measurement
fails due to a ``Buridan's ass''\ effect. The probability distribution for the
magnetization then\ develops not one but two narrow peaks at the
ferromagnetic values. During its evolution it goes through wide shapes
extending between these values.

\end{abstract}

\maketitle

%%%%%%%%%%%%%%%%%%%%%%%%%%%%%%%%%%%%%%%%%%%%
%% MAINMATTER
%%%%%%%%%%%%%%%%%%%%%%%%%%%%%%%%%%%%%%%%%%%%

\section{Quantum measurements, a problem of statistical mechanics}

A quantum measurement has well known features which relate the initial state
of the tested system $\mathrm{S}$ and the final state of the compound system
$\mathrm{S}+\mathrm{A}$ including \textrm{S} and the apparatus \textrm{A}~[1].
Von Neumann's reduction of the state postulates that an ideal measurement erases
the off-diagonal blocks of the density matrix of $\mathrm{S}+\mathrm{A}$ in a
basis where the measured observable is diagonal. Born's rule provides the
probability of observing on the apparatus such or such value, and this value
is correlated with the state of $\mathrm{S}$ after measurement. However, in
order to understand how these features arise in an actual experiment, we need
to analyze the dynamical process undergone by the coupled system
$\mathrm{S}+\mathrm{A}$, to dig out the time scales involved, and to exhibit
the specific properties of the quantum system $\mathrm{A}$ required so that it
can be used as an apparatus.

At the initial time of the measurement, $\mathrm{S}$ and $\mathrm{A}$ are
uncorrelated, with density operators $\hat{r}\left(  0\right)  $ and
$\mathcal{\hat{R}}\left(  0\right)  $, respectively. The full density operator
$\mathcal{\hat{D}}\left(  t\right)  $, initially equal to $\mathcal{\hat{D}%
}\left(  0\right)  =\hat{r}\left(  0\right)  \otimes\mathcal{\hat{R}}\left(
0\right)  $, evolves according to the Liouville--von Neumann equation. We wish
to explain how it eventually reaches at the final time $t_{\mathrm{f}}$ the
form
\begin{equation}
\mathcal{\hat{D}}\left(  t_{\mathrm{f}}\right)  =\sum_{i}\left(  \hat{\Pi}%
_{i}\hat{r}\left(  0\right)  \hat{\Pi}_{i}\right)  \otimes\mathcal{\hat{R}%
}_{i}\label{001}%
\end{equation}
which embeds the standard properties of ideal measurements recalled above. We
denote by $\hat{\Pi}_{i}$ the projection operator over the eigenspace of the
measured observable $\hat{A}$ labelled by its eigenvalue $A_{i}$, and by
$\mathcal{\hat{R}}_{i}$ the set of possible final states of the apparatus
$\mathrm{A}$. Each one, $\mathcal{\hat{R}}_{i}$, is characterized by some
value of a pointer variable of $\mathrm{A}$, in one-to-one correspondence with
the value $A_{i}$ for $\mathrm{S}$, and which can be observed or registered.
The expression (\ref{001}) means that, in a set of repeated experiments, a
well defined outcome $i$ occurs simultaneously for $\mathrm{A} $ (on which it
can be observed) and for $\mathrm{S}$, which is thereby prepared in the state
$\hat{\Pi}_{i}\hat{r}\left(  0\right)  \hat{\Pi}_{i}/\operatorname*{Tr}%
_{\mathrm{S}}\hat{\Pi}_{i}\hat{r}\left(  0\right)  $ through selection of the
outcome $i$ for $\mathrm{A}$. Such events occur with a classical probability
$\operatorname*{Tr}_{\mathrm{S}}\hat{\Pi}_{i}\hat{r}\left(  0\right)  $.

The quantum system $\mathrm{S}$ is microscopic, and we wish its interaction
with $\mathrm{A}$ to be sufficiently weak so that the final state involves no
other correlation between $\mathrm{S}$ and $\mathrm{A}$ than the ones
exhibited in eq. (\ref{001}). The system $\mathrm{A}$ should therefore be able
to switch from $\mathcal{\hat{R}}\left(  0\right)  $ to any one of the states
$\mathcal{\hat{R}}_{i}$ under the effect of a very small perturbation. It is
thus subject to a \textit{bifurcation}. The state $\mathcal{\hat{R}}\left(
0\right)  $ of $\mathrm{A}$ should be metastable when the interaction with
$\mathrm{S}$ is not yet turned on, and it should evolve towards any one of the
states $\mathcal{\hat{R}}_{i}$ when triggered by this small interaction.
Moreover, a robust and permanent registration excludes the possibility of
transition from one state $\mathcal{\hat{R}}_{i}$ to another over a reasonable
delay. Such states should thus be stable, and should have no overlap with one
another, which imposes the apparatus $\mathrm{A}$ to be a macroscopic object.
There should also exist some symmetry between the possible pointer variables,
so that no bias is introduced by the apparatus. Otherwise the probability of
observing the result $i$ would not be $\operatorname*{Tr}_{\mathrm{S}}\hat
{\Pi}_{i}\hat{r}\left(  0\right)  $ as stated by Born's rule, but would also
depend on the lack of symmetry of $\mathrm{A}$.

A natural means for achieving these requirements is to choose $\mathrm{A}$ as
a macroscopic system which can undergo a phase transition with broken
invariance. In some way or another, many actual apparatuses rely on such a
transition (although bifurcations without broken symmetry are currently used).
The pointer variable is then identified with the order parameter, a quantity
that can be observed without perturbing the apparatus. The initial state
$\mathcal{\hat{R}}\left(  0\right)  $ is chosen as the symmetric state which
is metastable upon sudden cooling, and the states $\mathcal{\hat{R}}_{i}$ are
the possible stable states with broken symmetry. Whereas in statistical
mechanics the invariance is broken either spontaneously or under the effect of
a scalar source, in a quantum measurement the breaking is triggered by
coupling with $\mathrm{S}$. Although this coupling should be weak, its effects
on $\mathrm{A}$ are considerably amplified, due to the metastability of
$\mathcal{\hat{R}}\left(  0\right)  $. On the other hand, we wish the
measurement to perturb the microscopic system $\mathrm{S}$ as little as
possible in spite of the macroscopic nature of $\mathrm{A}$: we wish the
diagonal blocks $\hat{\Pi}_{i}\hat{r}\left(  0\right)  \hat{\Pi}_{i}$ to be
kept unchanged, while the laws of quantum mechanics should prevent the off-diagonal
blocks of $\hat{r}\left(  0\right)  $ to subsist. Thus, although $\mathrm{S}$
behaves as a source for the phase transition, its quantum nature is essential
for a full description of the dynamics.

\, From the large size of $\mathrm{A}$ and the small size of $\mathrm{S}$ we can
anticipate that this dynamics of the coupled system $\mathrm{S}+\mathrm{A}$
during the measurement involves several time scales. The variables associated
with $\mathrm{S}$ are expected to vary rapidly from the outset; during this
stage $\mathrm{A}$ hardly changes. The establishment of correlations and the
registration of the result in the form of a finite order parameter will
require a much longer delay, which is required to let the macroscopic system
$\mathrm{A}$ reach equilibrium.

\section{A solvable model}

We have worked out the above ideas on an exactly solvable model, where
$\mathrm{S}$ is a single spin $\frac{1}{2}$ represented by the Pauli operators
$\hat{s}_{x}$, $\hat{s}_{y}$, $\hat{s}_{z}$, and where $\mathrm{A}$ simulates
a magnetic dot containing $N$ spins
% TCIMACRO{\TeXButton{bold-hat-sigma}{\mbox{\boldmath{$\hat{\sigma}$}}}}%
% BeginExpansion
\mbox{\boldmath{$\hat{\sigma}$}}
% EndExpansion
$^{\left(  n\right)}$($n=1$, $2$, ...\ $N$) which can undergo an Ising phase
transition [2]. (For other models, see [3-8].)
The order parameter is the average magnetization along $z$,
\begin{equation}
\hat{m}=\frac{1}{N}\sum_{n=1}^{N}\hat{\sigma}_{z}^{\left(  n\right)}
\text{,}\label{002}%
\end{equation}
which takes at equilibrium the paramagnetic value $\left\langle \hat
{m}\right\rangle =0$ above the Curie temperature, or either one of the two
ferromagnetic values $\left\langle \hat{m}\right\rangle =\pm m_{\mathrm{F}}$
below. In both cases the statistical fluctuation is small as $1/\sqrt{N}$.
Thermal equilibrium of $\mathrm{A}=\mathrm{M}+\mathrm{B}$ is ensured by a weak
interaction of these magnetic degrees of freedom $\mathrm{M}$ with a thermal
bath $\mathrm{B}$, represented by a model describing the phonons in the dot.
The full Hamiltonian of $\mathrm{S}+\mathrm{A}$ has therefore the form
$\hat{H}=\hat{H}_{\mathrm{SA}}+\hat{H}_{\mathrm{A}}$, where
\begin{equation}
\hat{H}_{\mathrm{SA}}=-g\hat{s}_{z}\sum_{n=1}^{N}\hat{\sigma}_{z}^{\left(
n\right)  }=-Ng\hat{s}_{z}\hat{m}\label{003}%
\end{equation}
accounts for the interaction between the $z$-component of the spin
%TCIMACRO{\TeXButton{bold-hat-s}{\mbox{\boldmath{$\hat{s}$}}} }%
%BeginExpansion
\mbox{\boldmath{$\hat{s}$}}
%EndExpansion
and the apparatus, and where $\hat{H}_{\mathrm{A}}=\hat{H}_{\mathrm{M}}%
+\hat{H}_{\mathrm{MB}}+\hat{H}_{\mathrm{B}}$ includes the Ising interaction
\begin{equation}
\hat{H}_{\mathrm{M}}=-\frac{J}{2N}\sum_{n,n^{\prime}}\hat{\sigma}_{z}^{\left(
n\right)  }\hat{\sigma}_{z}^{\left(  n^{\prime}\right)  }=-\frac{NJ}{2}\hat
{m}^{2}\label{004}%
\end{equation}
between the spins of the apparatus (which all interact pairwise, unlike in our previous
studies, where quartets interact), the
coupling
\begin{equation}
\hat{H}_{\mathrm{MB}}=\sqrt{\gamma}\sum_{n=1}^{N}\sum_{a=x,y,z}\hat{\sigma
}_{a}^{\left(  n\right)  }\hat{B}_{a}^{\left(  n\right)  }\label{005}%
\end{equation}
of these spins with the bath through phonon operators $\hat{B}_{a}^{\left(
n\right)  }$, and the phonon Hamiltonian $\hat{H}_{\mathrm{B}}$. The initial
state $\mathcal{\hat{R}}\left(  0\right)  =\hat{R}_{\mathrm{M}}\left(
0\right)  \otimes\hat{R}_{\mathrm{B}}\left(  0\right)  $ of $\mathrm{A}$
factorizes into contributions of the magnet and of the bath, where $\hat
{R}_{\mathrm{M}}\left(  0\right)  =\hat{I}/2^{N}$ represents a completely
disordered paramagnetic state, prepared by bringing $\mathrm{M}$ at a high
temperature before the measurement, and where $\hat{R}_{\mathrm{B}}\left(
0\right)  \propto e^{-\hat{H}_{\mathrm{B}}/T}$ describes equilibrium of the
phonons at the temperature $T$. This temperature lies below the Curie
temperature ($T<J$) so that $\mathcal{\hat{R}}\left(  0\right)  $ is
metastable: it may transit towards the stable ferromagnetic states, in a
rather large time if the magnet-bath coupling $\gamma$ is weak. The bath
Hamiltonian $\hat{H}_{\mathrm{B}}$ will enter the problem only through the
autocorrelation function $\left\langle \hat{B}_{a}^{\left(  n\right)  }\left(
t\right)  \hat{B}_{a^{\prime}}^{\left(  n^{\prime}\right)  }\left(  t^{\prime
}\right)  \right\rangle $, which under rather general conditions has the form
\begin{equation}
\operatorname*{Tr}\hat{R}_{\mathrm{B}}\left(  0\right)  \hat{B}_{a}^{\left(
n\right)  }\left(  t\right)  \hat{B}_{a^{\prime}}^{\left(  n^{\prime}\right)
}\left(  t^{\prime}\right)  =\delta_{nn^{\prime}}\delta_{aa^{\prime}}K\left(
t-t^{\prime}\right)  \text{ ,}\label{006}%
\end{equation}
\begin{equation}
\tilde{K}\left(  \omega\right)  =\int_{-\infty}^{+\infty}dte^{-i\omega
t}K\left(  t\right)  =\frac{\hbar^{2}\omega}{4}\frac{e^{-\left|
\omega\right|  /\Gamma}}{e^{\hbar\omega/T}-1}\text{ .}\label{007}%
\end{equation}
We have denoted the Debye cutoff as $\Gamma$.

Since $\gamma$ is weak, the bath can be eliminated by means of a second-order
treatment of $\hat{H}_{\mathrm{MB}}$. From the Liouville--von Neumann equation
for $\mathcal{\hat{D}}$, we thus find, for the partial trace $\hat
{D}=\operatorname*{Tr}_{\mathrm{B}}\mathcal{\hat{D}}$ which describes the
joint evolution of $\mathrm{S}$ and $\mathrm{M}$ (in the presence of the
bath), the equation
\begin{equation}
\frac{d\hat{D}}{dt}=\frac{1}{i\hbar}\left[  \hat{H}_{\mathrm{SA}}+\hat
{H}_{\mathrm{M}},\hat{D}\right]  +\frac{\gamma}{\hbar^{2}}\sum_{n=1}^{N}%
\sum_{a=x,y,z}\int_{0}^{t}dt^{\prime}\left[  \hat{\sigma}_{a}^{\left(
n\right)  },\ \hat{D}\left(  t\right)  K\left(  -t^{\prime}\right)
\hat{\sigma}_{a}^{\left(  n\right)  }\left(  t^{\prime}\right)  -K\left(
t^{\prime}\right)  \hat{\sigma}_{a}^{\left(  n\right)  }\left(  t^{\prime
}\right)  \hat{D}\left(  t\right)  \right]  \text{ .}\label{008}%
\end{equation}
We have introduced, in the interaction representation,
\begin{equation}
\hat{\sigma}_{a}^{\left(  n\right)  }\left(  t\right)  \equiv\hat{U}\left(
t\right)  \hat{\sigma}_{a}^{\left(  n\right)  }\hat{U}^{\dag}\left(  t\right)
\text{ ,}\label{009}%
\end{equation}
\begin{equation}
\hat{U}\left(  t\right)  =\exp\left[  \left(  \hat{H}_{\mathrm{SA}}+\hat
{H}_{\mathrm{M}}\right)  t/i\hbar\right]  \text{ .}\label{010}%
\end{equation}

\section{Disappearance of Schr\"odinger cats}
%Reduction of the state}

During an ideal quantum measurement, the quantity to be measured should not
change. This is expressed here by the commutation of $\hat{H}$ with $\hat
{s}_{z}$, or equivalently with the projection operators $\hat{\Pi}_{i}$ on the
eigenstates $i=\uparrow$ or $\downarrow$ of $\hat{s}_{z}$. Hence, eq.
(\ref{008}) can be decomposed into four equations governing the blocks
$\hat{D}_{ij}\equiv\hat{\Pi}_{i}\hat{D}\hat{\Pi}_{j}\left(  i,j=\uparrow\text{
or }\downarrow\right)  $. Each $\hat{D}_{ij}$ is an operator in the space of
$\mathrm{M}$.

This decoupling allows us to treat separately the off-diagonal and diagonal
blocks. The evolution of $\hat{D}_{\uparrow\downarrow}=\hat{D}_{\downarrow
\uparrow}^{\dagger}$ governed by (\ref{008}) has been studied elsewhere
[2, 9]. We briefly recall the results here. During a very brief reduction time
$\tau_{\mathrm{red}}=\hbar/\sqrt{2N}g$, all elements of $\hat{D}%
_{\uparrow\downarrow}$, which describe correlations between the components
$\hat{s}_{x}$ or $\hat{s}_{y}$ of the spin $\mathrm{S}$ and the various spins
%TCIMACRO{\TeXButton{bold-hat-sigma}{\mbox{\boldmath{$\hat{\sigma}$}}}}%
%BeginExpansion
\mbox{\boldmath{$\hat{\sigma}$}}%
%EndExpansion
$^{\left(  n\right)  }$ of $\mathrm{A}$, decrease down to zero. The initial
order exhibited by the non-vanishing value of $\left\langle \hat{s}%
_{x}\right\rangle $ or $\left\langle \hat{s}_{y}\right\rangle $ is scattered
into a very large number of small correlations between $\hat{s}_{x}$ or
$\hat{s}_{y}$ and the $z$-component of the many spins of $\mathrm{A}$.

This relaxation process is governed by the interaction term $\hat H_{\rm SA}$.
If the Hamiltonian if ${\rm S+A}$ did reduce to this term, the lost order would surge back in
$\hat D_{\uparrow\downarrow}$ at the time $\pi\hbar/(2g)$, producing detectable
effects through a recurrence process.
Such recurrences are hindered, and irreversibility of the collapse is ensured, owing to the
presence of the spin-bath coupling term $\hat H_{\rm MB}$, provided
$\gamma \gg g^2 /N \hbar^2 \Gamma^2$. The initial order is then lost once and for all
into the bath degrees of freedom. In the language of NMR, this irretrievable loss of phase
coherence is similar to a spin-lattice relaxation process.
However, an alternative mechanism involving no phonon bath can also suppress the recurrences,
for an apparatus A consisting only of the magnet M. Indeed, all the matrix elements of
$\hat D_{\uparrow\downarrow}$ remain negligible at all times after  $\tau_{\rm red}$
if the system-magnet interaction $\hat H_{\rm SA}$ involves slightly different coupling
constants $g_{n}$ between $\hat{s}_{z}$ and the apparatus spins
$\hat{\sigma}_{z}^{\left(  n\right)  }$, such that their relative fluctuation satisfies
$\delta g/g\gg1/\sqrt{N}$.
This second mechanism is comparable to the relaxation in NMR due to inhomogeneity of the
external Larmor field. As in Hahn's spin echoes, one can imagine to retrieve here the initial
order associated with $\hat D_{\uparrow\downarrow}$ by means of an adequate setup of pulses.

    Altogether, the von Neumann collapse of the state, which can be regarded as the disappearance
of "Schrödinger cats" in a measurement, is explained as a relaxation process, which is rapid due
to the large size of $N$.
An important role is also played by the mixed nature of the initial state of our apparatus.
In most theoretical discussions of the measurement process a pure initial state is assumed.
This is not realistic, however, since it is impossible to control the macroscopic number of
degrees of freedom of an apparatus.
The relaxation related to the collapse should not be confused with standard decoherence.
It does not necessarily require the thermal bath, it depends on the measured observable,
and it takes place over a time $\tau_{\mathrm{red}}$ which involves $g$, not $T$.

\section{Registration by the apparatus}

We now focus on the evolution of the diagonal blocks $\hat{D}_{\uparrow
\uparrow}$ and $\hat{D}_{\downarrow\downarrow}$, which describe the
expectation values of $\hat{s}_{z}$ and of the spin operators of $\mathrm{M}$,
their fluctuations and all their correlations. This evolution should end up in
a final state of the form (\ref{001}), expressing the registration of the
measurement: a complete correlation established between the final value $\pm1$
of $\hat{s}_{z}$ and the final ferromagnetic equilibrium state, characterized
by the sign of the order parameter $\left\langle \hat{m}\right\rangle =\pm
m_{\mathrm{F}}$. The diagonal block $\hat{D}_{\uparrow\uparrow}$ is associated
with the occurrence of $s_{z}=+1$. We have to prove that in this block the
probability distribution for $\hat{m}$ tends for large times to become sharply
peaked around $+m_{\mathrm{F}}$, without any contribution from the region of
$-m_{\mathrm{F}}$. In the equation of motion for $\hat{D}_{\uparrow\uparrow}$,
obtained from (\ref{008}), the operator $\hat{H}_{\mathrm{SA}}+\hat
{H}_{\mathrm{M}}$ depends only on the eigenvalue of $\hat{s}_{z}$ in this sector,
equal here to $+1$, and on
the observable $\hat{m}$, the successive eigenvalues of which are separated by
a distance $2/N$. This introduces through (\ref{003}), (\ref{004}) and
(\ref{009}), (\ref{010}) the operators
\begin{equation}
\hbar\hat{\Omega}_{\pm}=-Ng\left[  \left(  \hat{m}\pm\frac{2}{N}\right)
-\hat{m}\right]  -\frac{NJ}{2}\left[  \left(  \hat{m}\pm\frac{2}{N}\right)
^{2}-\hat{m}^{2}\right]  =\mp2\left(  g+J\hat{m}\right)  -2J/N\equiv
\mp2h\left(  \hat{m}\right)  -2J/N\text{ ,}\label{011}%
\end{equation}
where $h\left(  \hat{m}\right)= g+J\hat{m}$ behaves as an operator-valued
self-consistent field. We also introduce the function
\begin{equation}
\tilde{K}_{t}\left(  \omega\right)  \equiv\int_{-t}^{+t}dse^{-i\omega
s}K\left(  s\right)  =\int_{-\infty}^{+\infty}d\omega^{\prime}\tilde{K}\left(
\omega^{\prime}\right)  \frac{\sin\left(  \omega^{\prime}-\omega\right)
t}{\pi\left(  \omega^{\prime}-\omega\right)  }\text{ ,}\label{012}%
\end{equation}
in which $\omega$ can be replaced by the operators $\hat{\Omega}_{\pm}$, and
thus find the reduced equation of motion
\begin{equation}
\frac{d\hat{D}_{\uparrow\uparrow}}{dt}=\frac{2\gamma}{\hbar^{2}}\sum_{n=1}%
^{N}\left[  \hat{\sigma}_{-}^{\left(  n\right)  },\ \hat{D}_{\uparrow\uparrow
}\tilde{K}_{t}\left(  \hat{\Omega}_{-}\right)  \hat{\sigma}_{+}^{\left(
n\right)  }-\hat{\sigma}_{+}^{\left(  n\right)  }\tilde{K}_{t}\left(
\hat{\Omega}_{+}\right)  \hat{D}_{\uparrow\uparrow}\right]  \text{
.}\label{013}%
\end{equation}

This evolution conserves the trace $\operatorname*{Tr}_{\mathrm{M}}\hat
{D}_{\uparrow\uparrow}=\operatorname*{Tr}_{\mathrm{S}}\hat{\Pi}_{\uparrow}%
\hat{r}\left(  0\right)=r_{\uparrow\uparrow}(0)$, a normalization consistent with Born's rule.
Moreover the operator $\hat{D}_{\uparrow\uparrow}$ in the space of
$\mathrm{M}$ turns out to be simply a function of $\hat{m}$ at each time,
$\hat{D}_{\uparrow\uparrow}\left(  t\right)  =\Delta_{\uparrow\uparrow}\left(
\hat{m},t\right)  $, since this property holds at the initial time and is
preserved by the motion (\ref{013}). In fact, the knowledge of $\Delta
_{\uparrow\uparrow}\left(  \hat{m},t\right)  $ is equivalent to that of the
conditional probability $P_{\rm d}\left(  m,t\right)  $ for $\hat{m}$ to take the
discrete values $m=-1$, $-1+2/N$, ..., $1-2/N$, $1$ if $s_{z}$ equals $+1$,
which is expressed by
\begin{equation}
P_{\rm d}\left(  m,t\right)  =\frac{N!}{\left[  \frac{1}{2}N\left(  1+m\right)
\right]  !\left[  \frac{1}{2}N\left(  1-m\right)  \right]  !}\frac
{\Delta_{\uparrow\uparrow}\left(  m,t\right)  }{\operatorname*{Tr}\hat
{D}_{\uparrow\uparrow}}\text{ .}\label{014}%
\end{equation}
   From (\ref{013}) we find the equation of motion for $P_{\rm d}\left(  m,t\right)  $,
\begin{align}
\frac{\partial P_{\rm d}\left(  m,t\right)  }{\partial t} &  =\frac{\gamma N}%
{\hbar^{2}}\left[  \tilde{K}_{t}\left(  -\Omega_{+}\right)  \left(
1+m+\frac{2}{N}\right)  P_{\rm d}\left(  m+\frac{2}{N},t\right)  -\tilde{K}_{t}\left(
\Omega_{+}\right)  \left(  1-m\right)  P_{\rm d}\left(  m,t\right)  \right.
\label{015}\\
&  \left.  +\tilde{K}_{t}\left(  -\Omega_{-}\right)  \left(  1-m+\frac{2}%
{N}\right)  P_{\rm d}\left(  m-\frac{2}{N},t\right)  -\tilde{K}_{t}\left(  \Omega
_{-}\right)  \left(  1+m\right)  P_{\rm d}\left(  m,t\right)  \right]  \text{
,}\nonumber
\end{align}
where $\Omega_{\pm}$ are functions of $m$ defined by eq.(\ref{011}). For
shorthand we have denoted by $P_{\rm d}\left(  m,t\right)  $ instead of $P_{\uparrow
}\left(  m,t\right)  $ the conditional probability of $m$ associated with the
value $s_{z}=+1$ of the measured observable of $\mathrm{S}$, while the subscript `d'
indicates that $m$ is discrete here. We can introduce
likewise for the sector $\downarrow\downarrow$ the conditional probability
$P_{\downarrow}\left(  m,t\right)  $ associated with $s_{z}=-1$, which is
obtained by merely changing $g$ into $-g$ in the expression (\ref{011}) of
$\Omega_{\pm}$.

The registration process of the measurement is therefore just the same, in the
sector $\uparrow\uparrow$, as the relaxation of the Ising model towards
equilibrium under the influence of a field $+g$ and of the phonon bath, a
problem that we now study. The dynamical equation (\ref{015}) has been derived
from the Hamiltonian $\hat{H}_{\mathrm{A}}-Ng\hat{m}$ without any other
approximation than a weak magnet-bath coupling $\gamma\ll1$. The bath occurs
through $\tilde{K}_{t}$ defined by (\ref{007}), (\ref{012}), while the
Hamiltonian of $\mathrm{M}$ (including the field $+g$) occurs through the
energy shifts $\hbar\Omega_{\pm}$ given by (\ref{011}).

For sufficiently small $\gamma$ the evolution is slow and its time scale is
large compared to $\hbar/T$, so that $\tilde{K}_{t}\left(  \omega\right)  $
can be replaced in (\ref{015}) by $\tilde{K}\left(  \omega\right)  $. In such
a short-memory approximation, we have $\tilde{K}\left(  -\omega\right)
=\tilde{K}\left(  \omega\right)  e^{\hbar\omega/T}$, and the equation
(\ref{015}) for $P_{\rm d}\left(  m,t\right)  $ can be identified with a balance
equation, where the probability of each spin flip induced by the coupling with
the bath is given by Fermi's golden rule, and which might have been written
directly on phenomenological grounds. In terms of the entropy $S$ and the
energy $U$ of $\mathrm{M}$ associated with the density operator $\hat
{D}_{\uparrow\uparrow}$, eq. (\ref{015}) satisfies in this regime an
$H$-theorem
\begin{gather}
\frac{d}{dt}\left(  S-\frac{U}{T}\right)  =\frac{\gamma N}{\hbar^{2}}\sum
_{m}\tilde{K}\left(  \Omega_{-}\right)  \left[  e^{\hbar\Omega_{-}/T}\left(
1-m+\frac{2}{N}\right)  P_{\rm d}\left(  m-\frac{2}{N}\right)  -\left(  1+m\right)
P_{\rm d}\left(  m\right)  \right] \label{016}\\
\ln\frac{e^{\hbar\Omega_{-}/T}\left(  1-m+\frac{2}{N}\right)  P_{\rm d}\left(
m-\frac{2}{N}\right)  }{\left(  1+m\right)  P_{\rm d}\left(  m\right)  }\geq0\text{
,}\nonumber
\end{gather}
which implies that $\hat{D}_{\uparrow\uparrow}$ tends within normalization to
the equilibrium distribution $\exp N\left(  g\hat{m}+\frac{1}{2}J\hat{m}%
^{2}\right)  /T$, with possible invariance breaking for small $g$.

\section{Dynamics of the phase transition}

Our purpose is to study the dynamics of this relaxation, which for $T<J$
involves a bifurcation from the vicinity of $m=0$ towards that of $\pm
m_{\mathrm{F}}$. In the large $N$ limit, we may treat $m$ as a continuous
variable, $P\left(  m,t\right)=(N/2)P_{\rm d}\left(  m,t\right)$ being now normalized as
$\int_{-1}^{+1}dm\,P\left(  m,t\right)  =1$. From (\ref{007}), (\ref{011}) and (\ref{015})
we get for $t\gg\hbar/T$, keeping the terms of order $1$ and $1/N$,
\begin{equation}
\frac{\partial P\left(  m,t\right)  }{\partial t}=\frac{\partial}{\partial
m}\left[  -v\left(  m\right)  P\left(  m,t\right)  +\frac{1}{N}w\left(
m\right)  \frac{\partial P\left(  m,t\right)  }{\partial m}\right]  \text{
,}\label{017}%
\end{equation}
where, with now  $h(m)=g+Jm$ just involving the $c$-number $m$,
\begin{equation}
v\left(  m\right)  \equiv \frac{\gamma\, h(m)}{\hbar}\left(  1-m\,\coth\frac{h(m)}%
{T}+\frac{1}{N}\right)  \text{ ,}\label{018}%
\end{equation}
\begin{equation}
w\left(  m\right)  \equiv \frac{\gamma\, h(m)}{\hbar}\left(  \coth\frac{h(m)}%
{T}-m\right)  \text{ .}\label{019}%
\end{equation}

This type of equation has been extensively studied [10, 11]. We analyse below
its solution in the present context. The term in $1/N$ of (\ref{017}) is
negligible for smooth probabilities, but not for sharply peaked functions
$P\left(  m\right)  $, which occur at least in the initial state and in the
final equilibrium ferromagnetic states. In fact, this term is dominant in the
vicinity of the points where $m=\tanh h/T$ for which $v\left(  m\right)  $
vanishes. In particular, as implied by the inequality (\ref{016}), $P\left(
m,t\right)  $ tends for large times to an equilibrium shape characterized by
the vanishing of the square bracket in (\ref{017}). This condition entails
\begin{equation}
P\left(  m,\infty\right)  \propto\exp\left[  -\frac{N}{2}\left(
\frac{m-m_{\mathrm{F}}}{\delta_{\mathrm{F}}}\right)  ^{2}\right]  \text{
,}\label{020}%
\end{equation}
where $m_{\mathrm{F}}$ and $\delta_{\mathrm{F}}$ are given by
\begin{equation}
m_{\mathrm{F}}\left(  1-\frac{1}{N}\right)  =\tanh\frac{g+Jm_{\mathrm{F}}}%
{T}\text{\quad,\quad}m_{\mathrm{F}}\gtrless0\text{ ,}\label{021}%
\end{equation}
\begin{equation}
\frac{1}{\delta_{\mathrm{F}}^{2}}=\frac{1}{w}\frac{dv}{dm}%
\Bigg |_{m_{\mathrm{F}}}=\frac{1}{1-m_{\mathrm{F}}^{2}}-\frac{J}{T}>0\text{
.}\label{022}%
\end{equation}
As readily checked, we recover dynamically the expected canonical
distribution, including corrections for finite but large $N$. However, for
$g\ll Jm_{\mathrm{F}}^{3}$, eq. (\ref{021}) has two solutions such that
(\ref{022}) is positive, with nearly opposite values $\pm m_{\mathrm{F}}$.
Thus the asymptotic limit of $P\left(  m,t\right)  $ for large $t$ is not a
priori known: it is a linear combination of the two ferromagnetic
distributions (\ref{020}) peaked around $+m_{\mathrm{F}}$ and $-m_{\mathrm{F}%
}$. The present dynamical approach is necessary to explain how the evolution
towards equilibrium produces the breaking of invariance. In fact, the weights
$\mathcal{P}_{+}$\ and $\mathcal{P}_{-}$\ of the two terms $+m_{\mathrm{F}}$
and $-m_{\mathrm{F}}$ can only be found by solving the dynamical equation
(\ref{017}), which will allow us to express them in terms of the initial
condition $P\left(  m,0\right)  $. In the measurement problem, it is essential
that in the $\uparrow\uparrow$ sector, for $g>0$, the probability $P\left(
m,t\right)  $ concentrates for large times only around the positive
ferromagnetic value $+m_{\mathrm{F}}$; a finite weight $\mathcal{P}_{-}$ for
the peak $-m_{\mathrm{F}}$ would mean a wrong indication of the apparatus for
the system in the state $s_{z}=+1$. Moreover, we need the lifetime of the
initial paramagnetic state to be sufficiently large so that the interaction
between \textrm{S} and \textrm{A} can be turned on while $\mathrm{M}$ still
lies in the metastable paramagnetic state. In the theory of phase transitions,
we wish to find through which intermediate shapes $P\left(  m,t\right)  $
passes when the magnet $\mathrm{M}$ relaxes from the initial state to either
one of the ferromagnetic equilibrium states, and to determine the
probabilities of both occurrences.

Our equation (\ref{017}), which governs the time-dependence of the order
parameter, treated as a random variable, has the same form as the
Fokker--Planck equation for a Brownian particle in one dimension submitted to
an external influence. Its first term represents a deterministic drift, its
second term a diffusion process which tends to widen the distribution
$P\left(  m,t\right)  $. Let us first drop this diffusion term. Eq.
(\ref{017}) has then elementary solutions of the form $\delta\left(
m-m\left(  \mu,t\right)  \right)  $, where $m\left(  \mu,t\right)  $ is the
trajectory of a particle with initial position $m=\mu$ and
(position-dependent) velocity $dm/dt=v\left(  m\right)  $. (For very short
times, which are not relevant below, we find from (\ref{012}), (\ref{015})
that $v$ also depends on time, as $v\sim-\pi^{-1}\gamma\Gamma^{2}tm$, the sign
of which implies that the fixed point $m=0$ is initially stable.)\ The
functions $m\left(  \mu,t\right)$ is obtained by inverting the equation
\begin{equation}
t=\int_{\mu}^m %{m\left(  \mu,t\right)  }
\frac{dm'}{v\left(  m'\right)  }=\frac{\hbar}{\gamma}\int_{\mu}^m %{m\left(  \mu,t\right)  }
\frac{dm'}{h(m')\left[  1-m'\coth h(m')/T\right]
}\text{ .}\label{023}%
\end{equation}
This motion of $m$\ has three fixed points, the zeroes of $v\left(  m\right)
$. The closest to the origin, which for $g\ll Jm_{\mathrm{F}}^{3}$ lies at
\begin{equation}
m_{\mathrm{P}}=-\frac{g}{J-T}\text{ ,}\label{024}%
\end{equation}
is repulsive: if $\mu$ differs slightly from $m_{\mathrm{P}}$, the point
$m\left(  \mu,t\right)  $ moves astray. The other two fixed points are
attractors associated with the ferromagnetic phases near $+m_{\mathrm{F}}$ and
$-m_{\mathrm{F}}$. Thus, if the diffusion term is discarded, the initial
probability $P\left(  m,0\right)  $ is split into two parts $m>m_{\mathrm{P}}$
and $m<m_{\mathrm{P}}$, which will eventually result into sharp peaks located
at $+m_{\mathrm{F}}$ and $-m_{\mathrm{F}}$, respectively. In the evolution of
$P\left(  m,t\right)  $, the weight $P\left(  m,t\right)  dm$ is conserved
along the motion (\ref{023}). Hence, denoting by $\mu\left(  m,t\right)  $ the
inverse mapping of $m\left(  \mu,t\right)  $, the solution of eq. (\ref{017})
without the diffusion term is [11]
\begin{equation}
P\left(  m,t\right)  =P\left[  \mu\left(  m,t\right)  ,0\right]
\frac{ v\left[\mu\left(  m,t\right)  \right]  }{v\left(  m\right)}  \text{ ,}\label{025}%
\end{equation}
since $dm/d\mu=v\left(  m\right)  /v\left(  \mu\right)  $\ for given $t$.
We denoted $v(\mu(m,t))$ as $v[\mu(m,t)]$ and
$P(\mu\left(  m,t\right),0) $ as $P\left[  \mu\left(  m,t\right)  ,0\right] $.

However, as we already noted, diffusion is essential in the first stage of the
motion, when $P\left(  m,0\right)  $ is concentrated around small values of
$m$ where $v\left(  m\right)  $ is small. In this region, it is possible to
solve the full equation (\ref{017}) by expanding $v\left(  m\right)\approx (\gamma/\hbar)[g+(J-T)m]$
and $w\left(  m\right)\approx \gamma T/\hbar  $ in powers of $m$, leading to
\begin{equation}
\frac{\hbar}{\gamma}\frac{\partial P}{\partial t}=-\frac{\partial}{\partial
m}\left\{  \left[  g+\left(  J-T\right)  m\right]  P\right\}  +\frac{T}{N}%
\frac{\partial^{2}P}{\partial m^{2}}\text{ .}\label{026}%
\end{equation}
We can then find the explicit solution of (\ref{023}), and invert the mapping
$m\left(  \mu,t\right)  $,
\begin{equation}
m(\mu,t)=\mu\,e^{t/\theta}+\frac{g}{J-T}\left(e^{t/\theta}-1\right), \qquad
\mu\left(  m,t\right)  =me^{-t/\theta}-\frac{g}{J-T}\left(  1-e^{-t/\theta
}\right)  \text{ ,}\label{027}%
\end{equation}
where we introduced the time-scale
\begin{equation}
\theta\equiv\frac{\hbar}{\gamma\left(  J-T\right)  }\text{ .}\label{028}%
\end{equation}
The solution of (\ref{026}), found by means of a Fourier transform on $m$,
takes after some calculations the form
\begin{equation}
P\left(  m,t\right)  =\sqrt{\frac{N}{2\pi C}}\int d\xi e^{-N\xi^{2}%
/2C}P\left[  \mu\left(  m,t\right)  +\xi,0\right]  \frac{ v\left[
\mu\left(  m,t\right)  \right]}{v\left(  m\right)  }\text{ ,}\label{029}%
\end{equation}
where
\begin{equation}
C\equiv\left(  1-e^{-2t/\theta}\right) \frac{ T}{ J-T} \text{.}\label{030}%
\end{equation}
The result (\ref{029}) encompasses the drift induced by the first term of
(\ref{026}) and the diffusion induced by its second term. It turns out that
the cumulated effect of diffusion is equivalent to a blurring of the initial
value from which $m$ is issued in the deterministic motion (\ref{027}), over a
width $\sqrt{C}$ which increases with $t$. We note however that, for
$e^{t/\theta}\gg1$ the effect of diffusion remains constant, as $C$ tends then
to $C_{\infty}=T/\left(  J-T\right)  $.

Since, according to (\ref{027}), $\theta$ is also the characteristic time over
which $m\left(  \mu,t\right)  $ exponentially diverges from its initial value
$\mu$, we can match the effect (\ref{029}) of diffusion with the effect
(\ref{025}) of drift, even in regions where $m$ is no longer small, by writing
the solution of (\ref{017}) in the form (\ref{029}) where $\mu\left(
m,t\right)  $ is now obtained from (\ref{023}) rather than from (\ref{027}).
This result holds, up to the very large times when the expression (\ref{029})
gets concentrated near $\pm m_{\mathrm{F}}$. In the later stage of the
evolution, that we do not need to consider, the diffusion term becomes again
effective, and it determines the shape and width of the final probability
according to (\ref{020})-(\ref{022}). In order to discuss the behavior of
(\ref{029}) we note that when $J-T\ll J$, $m_{\mathrm{F}}^{2}\sim3\left(
J-T\right)  /J$, one has $m_{\mathrm{P}}\sim-g/\left(  J-T\right)  \ll m_{\mathrm{F}}%
$, while $v\left(  m\right)  $ has the form
\begin{equation}
v\left(  m\right)  \approx\frac{\left(  m-m_{\mathrm{P}}\right)  }{\theta
}\frac{m_{\mathrm{F}}^{2}-m^{2}}{m_{\mathrm{F}}^{2}}\text{ .}\label{031}%
\end{equation}
More generally, we note that (\ref{031}) has for $g\ll Jm_{\mathrm{F}}^{3}$
but arbitrary $T<J$ the same features as (\ref{018}): same zeroes, same
behavior for small $m$. For this qualitatively good model of $v\left(
m\right)  $ we can integrate explicitly (\ref{023}) for $\left|
m_{\mathrm{P}}\right|  \ll m_{\mathrm{F}}$ as
\begin{equation}
e^{t/\theta}=\frac{m-m_{\mathrm{P}}}{\mu-m_{\mathrm{P}}}\sqrt{\frac
{m_{\mathrm{F}}^{2}-\mu^{2}}{m_{\mathrm{F}}^{2}-m^{2}}}\text{ ,}\label{032}%
\end{equation}
or equivalently
\begin{equation}
m\left(  \mu,t\right)  =\frac{\left[  \mu e^{t/\theta}-m_{\mathrm{P}}\left(
e^{t/\theta}-1\right)  \right]  m_{\mathrm{F}}}{\sqrt{m_{\mathrm{F}}%
^{2}+\left(  \mu-m_{\mathrm{P}}\right)  ^{2}\left(  e^{2t/\theta}-1\right)  }%
}\text{ ,}\label{033}%
\end{equation}
\begin{equation}
\mu\left(  m,t\right)  =m_{\mathrm{P}}+\frac{\left(  m-m_{\mathrm{P}}\right)
e^{-t/\theta}m_{\mathrm{F}}}{\sqrt{m_{\mathrm{F}}^{2}-m^{2}\left(
1-e^{-2t/\theta}\right)  }}\text{ .}\label{034}%
\end{equation}
We take as initial state the narrow distribution
\begin{equation}
P\left(  m,0\right)  =\sqrt{\frac{N}{2\pi\delta_{0}^{2}}}\exp\left[  -\frac
{N}{2}\left(  \frac{m-m_{0}}{\delta_{0}}\right)  ^{2}\right]  \text{
,}\label{035}%
\end{equation}
where $m_{0}$ characterizes a possible deviation from the paramagnetic state
(for which $m_{0}=0$). The width $\delta_{0}$, determined by the initial
equilibrium of $\mathrm{M}$ at a temperature $T_{0}$ higher than $J$, is given
by $\delta_{0}^{2}=T_{0}/\left(  T_{0}-J\right)$.
When quenching from $T_0=\infty$, as is done in the figures,
one has $m_0=0$ and $\delta_0=1$.
Use of (\ref{032}) in (\ref{029}) and integration yield
\begin{equation}
P\left(  m,t\right)  =\sqrt{\frac{N}{2\pi\left(  C+\delta_{0}^{2}\right)  }%
}\exp\left[  -\frac{N}{2}\frac{\left(  \mu-m_{0}\right)  ^{2}}{C+\delta
_{0}^{2}}\right]  \frac{ v\left[
\mu\left(  m,t\right)  \right] }{v\left(  m\right)  }\text{
,}\label{036}%
\end{equation}
where $C$ is given by (\ref{030}), $\mu$ is the function of $m$ and $t$ given
by (\ref{034}), and $\partial\mu/\partial m=v\left(  \mu\right)  /v\left(
m\right)  $.

The expression (\ref{036}) holds at all times, except in the very last stage
of the evolution, when $P\left(  m,t\right)  $, already concentrated near
$+m_{\mathrm{F}}$ and $-m_{\mathrm{F}}$, switches into the shape (\ref{020}).
We can use it, however, to evaluate the probability $\mathcal{P}_{+}$ or
$\mathcal{P}_{-}$ that the initial state (\ref{035}) ends up at one or the
other ferromagnetic states $+m_{\mathrm{F}}$ or $-m_{\mathrm{F}}$. Indeed, in
the mapping $m\left(  \mu,t\right)  $, the points $\mu$ associated with the
vicinity of the final point $m=+m_{\mathrm{F}}$ are those for which
$\mu>m_{\mathrm{P}}$. Taking thus $\mu$ instead of $m$ as an integration
variable in (\ref{036}) and letting $t\rightarrow\infty$, we find
\begin{equation}
\mathcal{P}_{-}=\frac{1}{2}\operatorname{erfc}\left[  \sqrt{\frac{N}{2}}%
\frac{b}{\delta}\right]  \text{ ,}\label{037}%
\end{equation}
where
\begin{equation}
b\equiv-m_{\mathrm{P}}+m_{0}=\frac{g}{J-T}+m_{0}\label{038}%
\end{equation}
measures the bias due to the external field $g$ and to the shift $m_{0}$ in
the initial distribution, where
\begin{equation}
\delta^{2}\equiv C_{\infty}+\delta_{0}^{2}=\frac{T}{J-T}+\delta_{0}%
^{2}\label{039}%
\end{equation}
accounts for the modification of the initial width $\delta_{0}/\sqrt{N}$ due
to diffusion, and where
\begin{equation}
\operatorname{erfc}\left(  x\right)  \equiv\frac{2}{\sqrt{\pi}}\int
_{x}^{\infty}e^{-t^{2}}dt\text{ .}\label{040}%
\end{equation}

\section{Unique final state}

We can therefore distinguish two regimes which display several qualitatively
different features. In the first case, the argument of $\operatorname{erfc}$
in (\ref{037}) is large, so that the magnet $\mathrm{M}$ nearly certainly reaches the
ferromagnetic state $+m_{\mathrm{F}}$: the bifurcation is ineffective. This
occurs, for an unbiased initial state ($m_{0}=0$), when
\begin{equation}
g\gg\frac{\left(  J-T\right)  \delta}{\sqrt{N}}\text{ ;}\label{041}%
\end{equation}
this also occurs for $g=0$ when the shift $m_{0}$ of the initial state with
respect to paramagnetism satisfies
\begin{equation}
m_{0}\gg\frac{\delta}{\sqrt{N}}\text{ .}\label{042}%
\end{equation}
In this situation, the duration of the relaxation process is of order
$\theta=\hbar/\gamma\left(  J-T\right)  $. More precisely, if we consider that
ferromagnetism is reached when the distribution $P\left(  m,t\right)  $ is
peaked near $m=0.95$ $m_{\mathrm{F}}$, this time, the registration time for a
measurement, is found from (\ref{032}) to be
\begin{equation}
\tau_{\mathrm{reg}}=\frac{\hbar}{\gamma\left(  J-T\right)  }\ln\frac
{3m_{\mathrm{F}}}{b}\text{ .}\label{043}%
\end{equation}

Since the range $t<\tau_{\mathrm{reg}}$ does not involve $N$, it is seen from
(\ref{034}) that $N$ occurs in the expression (\ref{036}) of $P\left(
m,t\right)  $ only as a factor in the exponent. Hence $P\left(  m,t\right)  $
displays at all times a single peak, see Fig. 1, that is narrow as $1/\sqrt{N}$
and located at the point $m$ where $\mu\left(  m,t\right)  $ equals $m_{0}$.
The position of this peak is thus given by (\ref{033}) in which $\mu$ is replaced
by $m_{0}$. Its width is seen from (\ref{036}) to equal
$\delta\left(  t\right)  /\sqrt{N}$, where
\begin{equation}
\delta\left(  t\right)  =\sqrt{C+\delta_{0}^{2}}\ \frac{v\left(  m\right)
}{v\left(  \mu\right)  }\text{ .}\label{044}%
\end{equation}
When $m\left(  m_{0},t\right)  $ is still close to the origin, diffusion lets
$C$ increase up to $C_{\infty}=T/\left(  J-T\right)  $. The widening of the
distribution $P\left(  m,t\right)  $ is then governed by the dispersion of the
speed $v\left(  m\right)  $ of the drift motion [10]: the head of the
distribution, with larger $m$, proceeds faster than its tail. The maximum of
$\delta\left(  t\right)  $ is thus attained when the peak of $P\left(
m,t\right)  $ is located at the value of $m$ which lets $v\left(  m\right)  $
be maximum, that is, $m=m_{\mathrm{F}}/\sqrt{3}$, reached for
\begin{equation}
t=\frac{\hbar}{\gamma\left(  J-T\right)  }\ln\frac{m_{\mathrm{F}}}{\sqrt
{2}\ b}\text{ .}\label{045}%
\end{equation}
We have then
\begin{equation}
\delta_{\max}=\frac{2m_{\mathrm{F}}\delta}{3\sqrt{3}\ b}\text{ .}\label{046}%
\end{equation}
Later on, $\delta\left(  t\right)  $ decreases, again due to the gradient in
the drift velocity, which is now negative. This effect is, in the last stage,
counterbalanced by diffusion which prevents $\delta\left(  t\right)  $ from
decreasing below the equilibrium value $\delta_{\mathrm{F}}$ given by
(\ref{022}).

\begin{figure}
  \includegraphics[height=.25\textheight]{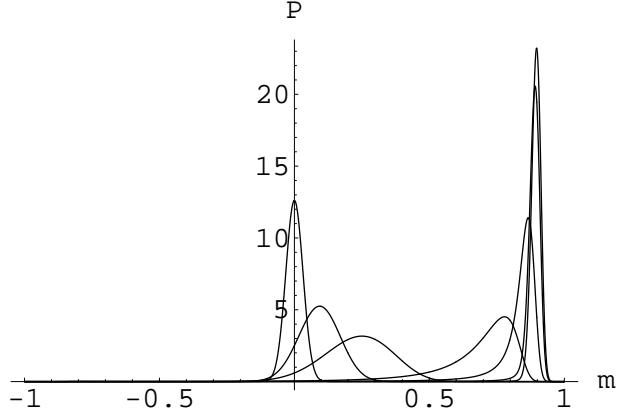}
  \caption{Transition from the initial paramagnetic state to the ferromagnetic state
with magnetization up, driven by the coupling $g$.
We have solved eq. (15) in the short-memory approximation, valid for $\gamma \ll 1$, where
$\tilde K_t(\omega)$ is replaced by (7), the Debye cutoff being irrelevant.
The parameters were chosen as $N=1000$, $T=0.65J$, $g=0.05J$, small but satisfying the condition (41).
The curves describe the probability distribution $P(m,t)$ at the times (from left to right)
$t/\theta=0$, $0.5$, $1$, $2.25$, $3$, $4$ and $5$; the latter curve is near the infinite time limit.
As initial distribution we took (35) with $m_0=0$ and $\delta_0=1$, corresponding to the paramagnet
$T_0=\infty$. The center of the peak moves from $m=0$ to $m_F=0.89707$  according to (23);
its width first increases up to (46) then decreases down to (22).
 }
\end{figure}

\section{Active bifurcation}

We now consider the second regime, for which the bifurcation is active in the
sense that neither $\mathcal{P}_{+}$ nor $\mathcal{P}_{-}$ vanish.
 Such a possibility for a macroscopic system to transit towards several (here two) possible
final states is known as spinodal decomposition.
This situation is a nuisance for the measurement problem, since it corresponds to a
failure in the establishment of correlations between the indication of
$\mathrm{A}$ and the residual state of $\mathrm{S}$. For the phase transition
problem, it corresponds to dynamics for which the external field $g$ and the
initial shift $m_{0}$ are too small to fully determine the phase towards which
the system $\mathrm{A}$ will relax. Since at large times $P\left(  m,t\right)
$ must have two sharp peaks, we know that some samples will reach
$+m_{\mathrm{F}}$, other ones $-m_{\mathrm{F}}$, but we are interested here in
the history of these samples. A natural question is the following: do the
samples that will end up at $+m_{\mathrm{F}}$ behave as above, with a
well-defined trajectory $m\left(  t\right)  $ within fluctuations of order
$1/\sqrt{N}$? More precisely, how does the distribution $P\left(  m,t\right)
$ evolve before it displays its final two-peaked equilibrium shape?

To answer this question, which arises when the conditions (\ref{041}),
(\ref{042}) are violated, we assume that the bias due to both the external
field and the initial condition is small for large $N$ as
\begin{equation}
b\equiv-m_{\mathrm{P}}+m_{0}=\frac{g}{J-T}+m_{0}=\sqrt{\frac{2}{N}}%
\ \lambda\delta\text{ ,}\label{047}%
\end{equation}
where $\delta$ is defined by (\ref{039}) and where $\lambda$ is either finite
or small. The probability distribution, given by (\ref{036}), begins to widen
under the effect of diffusion as
\begin{equation}
P\left(  m,t\right)  =\sqrt{\frac{N}{2\pi\left(  C+\delta_{0}^{2}\right)  }%
}\exp\left[  -\frac{N}{2}\frac{\left[  m-m_{0}+m_{\mathrm{P}}\left(
1-e^{-t/\theta}\right)  \right]  ^{2}}{C+\delta_{0}^{2}}\right]  \text{
,}\label{048}%
\end{equation}
where $C$ increases according to (\ref{030}). Even in case the initial
distribution (\ref{035}) extends only over values larger than $m_{\mathrm{P}}
$, that is, for $\delta_{0}\ll b=m_{0}-m_{\mathrm{P}}$, (\ref{048}) develops a
tail which may extend to the region $m<m_{\mathrm{P}}$, and which will give
rise to the non-vanishing probability (\ref{037}). This initial widening is
relayed by the widening due to the gradient of the drift velocities, which
will have here dramatic consequences. Indeed, whereas in the first regime the
whole relaxation process was achieved after a delay (\ref{043}) independent of
$N$, the probability is now not yet concentrated near $\pm m_{\mathrm{F}}$ for
times such that $e^{t/\theta}$ is of order $\sqrt{N}$, see Fig. 2.
In such a range of times, the distribution (\ref{036}) takes the form
\begin{equation}
P\left(  m,t\right)  =\frac{\alpha m_{\mathrm{F}}^{2}}{\sqrt{\pi}\ \left(
m_{\mathrm{F}}^{2}-m^{2}\right)  ^{3/2}}\exp\left[  -\left(  \frac{\alpha
m}{\sqrt{m_{\mathrm{F}}^{2}-m^{2}}}-\lambda\right)  ^{2}\right]  \text{
,}\label{049}%
\end{equation}
where we introduced Suzuki's scaling variable [11]
\begin{equation}
\alpha\equiv\sqrt{\frac{N}{2}}\ e^{-t/\theta}\frac{m_{\mathrm{F}}}{\delta
}\text{ .}\label{050}%
\end{equation}
The large $N$ limit is thus singular. During a long lapse of time, the
probability distribution (\ref{049}) presents no narrow peaks in $1/\sqrt{N}$,
although it behaves originally as (\ref{035}) and ends up as a sum of two
contributions (\ref{020}) around $+m_{\mathrm{F}}$ and $-m_{\mathrm{F}}$.
Instead, it is a smooth function, independent of $N$ and extending over a wide
range of values of $m$.

Let us illustrate this behavior by describing the evolution of (\ref{049}) in
the completely unbiased case $g=m_{0}=\lambda=0$, for which $P\left(
m,t\right)  $ is symmetric (Fig. 2). The problem is formally similar to the so-called
laser model [11] The initial peak (\ref{035}) with $m_{0}=0$ begins to widen
and to reach a finite width. It progressively covers most of the interval
$-m_{\mathrm{F}}$, $+m_{\mathrm{F}}$. In particular, at the time
\begin{equation}
t=\theta\ln\left(  \frac{m_{\mathrm{F}}}{\delta}\sqrt{\frac{N}{3}}\right)
\text{ ,}\label{051}%
\end{equation}
such that $\alpha^{2}=\frac{3}{2}$, the distribution
\begin{equation}
P\left(  m,t\right)  =\sqrt{\frac{3}{2\pi}}\frac{m_{\mathrm{F}}^{2}}{\left(
m_{\mathrm{F}}^{2}-m^{2}\right)  ^{3/2}}\exp\left(  -\frac{3}{2}\frac{m^{2}%
}{m_{\mathrm{F}}-m^{2}}\right) \label{052}%
\end{equation}
is very flat: near $m=0$, it behaves as $P\propto\exp\left(  -3m^{4}%
/4m_{\mathrm{F}}^{4}\right)  $; the ratio $P\left(  m,t\right)  /P\left(
0,t\right)  $ is still equal to $0.93$ for $m=0.5m_{\mathrm{F}}$, to $0.84$
for $m=0.6m_{\mathrm{F}}$, to $0.65$ for $m=0.7m_{\mathrm{F}}$. At later
times, the origin becomes a minimum and two maxima appear, lying at
\begin{equation}
m=\pm m_{\mathrm{F}}\sqrt{1-\frac{2\alpha^{2}}{3}}=\pm m_{\mathrm{F}}%
\sqrt{1-\frac{Ne^{-2t/\theta}m_{\mathrm{F}}^{2}}{3\delta^{2}}}\text{
.}\label{053}%
\end{equation}
These maxima move away from the origin towards $\pm m_{\mathrm{F}}$.
Originally not pronounced, they get narrower and narrower. For sufficiently
large times, when $\alpha$ becomes small, they turn into unsymmetrical peaks
near $m_{\mathrm{F}}$ (and $-m_{\mathrm{F}}$), having the shape
\begin{equation}
P\left(  m\right)  dm\sim\frac{e^{-1/x}dx}{2\sqrt{\pi}x^{3/2}}\text{\quad
,\quad}x=\frac{2}{\alpha^{2}}\frac{m_{\mathrm{F}}-m}{m_{\mathrm{F}}}\text{
,}\label{054}%
\end{equation}
with a maximum at $x=\frac{2}{3}$. Both peaks eventually reach the symmetric
equilibrium shape (\ref{020}) when the diffusion term dominates again. We
estimate the relaxation time by evaluating after which delay these two peaks
arrive at $\pm0.95m_{\mathrm{F}}$, which yields
\begin{equation}
\tau_{\mathrm{relax}}=\frac{\hbar}{\gamma\left(  J-T\right)  }\ln\left(
\frac{m_{\mathrm{F}}}{\delta}\sqrt{\frac{10N}{3}}\right)  \text{ .}\label{055}%
\end{equation}
This quantity is much larger than the relaxation times which occur in the
first regime, by a factor $\ln\sqrt{N}$. The relaxation time becomes even
infinite when $N\rightarrow\infty.$

For finite values of the bias, as measured by $\lambda$ defined by
(\ref{039}), (\ref{047}), the curve (\ref{049}) is no longer symmetric, but
the results are similar. For $\lambda>0$, the maximum initially at the origin
moves as $m\sim\lambda m_{\mathrm{F}}/\alpha$ while widening. At the time
(\ref{051}), there is still a single maximum, but the distribution has become
very wide, as for $\lambda=0$, with larger values for $m>0$ than for $m<0$. A
second maximum appears later on, at a negative value $m_{2}$ of $m$ given by
$m_{2}^{3}=-\lambda m_{\mathrm{F}}^{2}\sqrt{m_{\mathrm{F}}^{2}-2m_{2}^{2}%
}\ /\sqrt{6}$, and at a time such that $\alpha^{2}=\frac{3}{2}\left(
m_{\mathrm{F}}^{2}-m_{2}^{2}\right)  \left(  m_{\mathrm{F}}^{2}-2m_{2}%
^{2}\right)  /m_{\mathrm{F}}^{4}$. Here again the two maxima end up as narrow
peaks near $\pm m_{\mathrm{F}}$, but equilibrium is reached for
$+m_{\mathrm{F}}$ earlier than for $-m_{\mathrm{F}}$; the geometric mean of
these relaxation times is given by (\ref{055}). The main difference with the
symmetric case is the occurrence of unequal probabilities $\mathcal{P}_{+}$
and $\mathcal{P}_{-}$ for the final phases, given by (\ref{037}), which reads
\begin{equation}
\mathcal{P}_{\pm}=\frac{1}{2}\operatorname{erfc}\left(  \mp\lambda\right)
\text{ .}\label{056}%
\end{equation}
These values, determined by the dynamics, are expressed in terms of the
various parameters of the problem and of the initial conditions through
(\ref{039}), (\ref{047}).

\begin{figure}
  \includegraphics[height=.25\textheight]{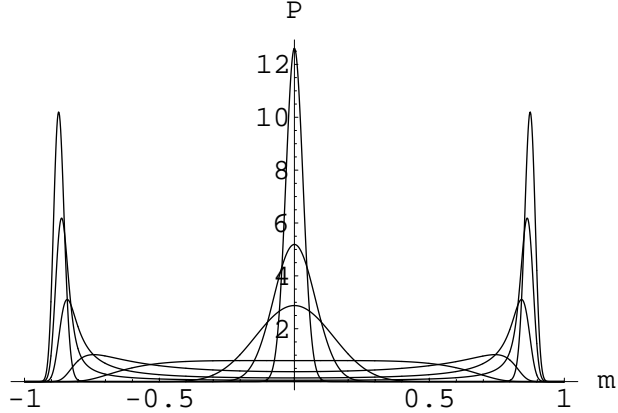}
  \caption{Transition from the initial paramagnetic state to the mixture of ferromagnetic states
% in absence of the coupling $g$
for the same situation as in Fig. 1, $N=1000$, $T/J=0.65$ but $g=0$.
The times are:
$t/\theta=0$, $0.5$, $1$, $2.25$, $3$, $4$, $5$ and $10$ (from top to bottom at $m=0$;
the last curve is very near the infinite-time limit).
% Transition...coupling g. The data are the same as in fig. 1 except for g=0, the successive times being now t/theta=.....
The distribution, initially peaked as $1/\sqrt{N}$, becomes flat in Suzuki's scaling regime (49)-(50)
(with $\lambda=0$ here). For large $t$, two peaks build up near $m_F=0.87206$ and $-m_F$,
with widths governed by the diffusion term $\sim 1/N$.
}
\end{figure}

\section{Buridan's ass effect}

Many among the properties exhibited above have been encountered in
different contexts long ago [10,11]. Indeed, the equation
(\ref{017}) which governs the dynamics of the present phase
transition has a very general form, describing directed Brownian
motion with an unstable fixed point. We have stressed the
existence for large $N$ of two contrasted regimes for the behavior
of the probability $P\left(  m,t\right)  $, depending on the
initial conditions and on the field. We propose to term the
second, anomalous situation ``Buridan's ass effect''. According to
an argument attributed to Jean Buridan, a dialectician of the
first half of the XIVth century, an ass placed equidistantly from
two identical bales of hay will stay there and starve to death
because it has no causal reason to choose one or the other [12].

Here, likewise, in a symmetric situation where the initial bifurcation may
lead the magnetization $m$ to $+m_{\mathrm{F}}$ or $-m_{\mathrm{F}}$ with
equal probabilities $\mathcal{P}_{+}=\mathcal{P}_{-}=\frac{1}{2}$, an infinite
time (\ref{055}) is required in the macroscopic limit $N\rightarrow\infty$ for
the system to relax. We find more generally the same huge duration of the
relaxation, from the paramagnetic state to either one of the ferromagnetic
states, when there is a bias which is sufficiently small so that neither one
of the probabilities $\mathcal{P}_{+}$ and $\mathcal{P}_{-}$ for reaching
$+m_{\mathrm{F}}$ and $-m_{\mathrm{F}}$ vanishes. In such a situation, if $N$
is large but finite, ``the ass'' $m$ will have reached one or the other bale
after the delay (\ref{055}), and we can predict with which probability: if we
perform the same experiment with many asses, we know which proportion will
have attained either side. However, for times $t<\theta$ the asses have
practically not yet moved, whereas for times such that $e^{t/\theta}$ is of
order $\sqrt{N}$, we can make no prediction: the asses are scattered on a wide
range between the two bales. This dispersion of the values of $m$ during a
long delay is a characteristic feature of the phenomenon.

In contrast, the situation is trivial for a larger bias. This may take place
either under the condition (\ref{041}) when the field $g$ is sufficiently
large (a wind which pushes the asses), or under the condition (\ref{042}) if
the initial state has a residual magnetization (the asses are significantly
closer to one of the bales). The process then takes a finite time (\ref{043})
when $N\rightarrow\infty$, and we can predict with little error (in
$1/\sqrt{N}$) where are the asses at each time.

The giant fluctuations which occur in Buridan's ass effect may be regarded as
a dynamical counterpart of those which occur at equilibrium near a critical
point. In both cases, the order parameter, although macroscopic, is not a
well-defined quantity even in the large $N$ limit, so that its behavior must
be described by means of statistical mechanics rather than standard
thermodynamics. In the present situation, no temperature can be defined for
the system $\mathrm{M}$, during the irreversible process which leads it from a
high-temperature equilibrium paramagnetic state (\ref{035}) to a ferromagnetic
equilibrium (\ref{020}) at the temperature $T$ of the bath, fixed below the
critical temperature $J$. However we may argue that in some sense $\mathrm{M}$
crosses its critical point during the dynamical process, in case it has some
probability to reach both phases $+m_{\mathrm{F}}$ and $-m_{\mathrm{F}}$ in
the end. The well-known critical fluctuations and critical slowing down thus
manifest themselves here by the large uncertainty about $m$ shown by $P\left(
m,t\right)  $ during the long delay (\ref{055}). Giant fluctuations in the
dynamics are also well-known for nucleation processes. However in the present
long-range model (\ref{004}) there is no space structure, so that we have only
to deal with the statistics at each time of the single order parameter $m$.

% \section{Application to the quantum measurement problem}

\section{CONCLUSION: BURIDAN'S ASS AND QUANTUM MEASUREMENT}

As regards the use of $\mathrm{A}=\mathrm{M}+\mathrm{B}$ as an apparatus to
measure the spin $\mathrm{S}$, two conditions should be satisfied. On the one
hand, the initial metastable paramagnetic state of $\mathrm{M}$, prepared by
setting its temperature at $T_{0}>J$, should have a long lifetime (\ref{055}),
so that the interaction (\ref{003}) between $\mathrm{S} $ and $\mathrm{A}$ can
be turned on before $\mathrm{M}$ has begun to relax. This is achieved provided
there is neither a significant external field $g_{0}$ nor a lack of symmetry
$m_{0}$ before the measurement, a condition expressed more precisely by
$\lambda\ll1$ in (\ref{047}), that is, using $\delta_{0}^{2}=T_{0}/\left(
T_{0}-J\right)  $,
\begin{equation}
\left(  \frac{g_{0}}{J-T}+m_{0}\right)  ^{2}\ll\frac{1}{N}\left(  \frac
{T}{J-T}+\frac{T_{0}}{T_{0}-J}\right)  \text{ .}\label{057}%
\end{equation}

On the other hand, the coupling $g$ between $\mathrm{S}$ and $\mathrm{A}$
should be sufficiently large so as to ensure faithfulness of the registration,
$+m_{\mathrm{F}}$ in the sector $\uparrow\uparrow$, $-m_{\mathrm{F}}$ in the
sector $\downarrow\downarrow$. This is achieved provided Buridan's ass effect
does not take place during the measurement process, that is, provided
\begin{equation}
\left(  \frac{g}{J-T}\right)  ^{2}\gg\frac{1}{N}\left(  \frac{T}{J-T}%
+\frac{T_{0}}{T_{0}-J}\right)  \text{ .}\label{058}%
\end{equation}
Then the registration time (\ref{043}) is finite for large $N$.

% \section{Conclusion}

We have chosen to study the model (\ref{004}) because its exact solution at
equilibrium can be obtained for large $N$ through a static mean-field
approach; we had the prejudice that its dynamics might also be solved exactly
through a time-dependent mean-field approach. It turns out that, although an
exact solution in the considered limiting case is available, it cannot reduce
to mean-field when Buridan's ass effect is present. A time-dependent
mean-field approach relies on the existence of trajectories $m\left(
t\right)  $ around which $m$ has negligible statistical fluctuations. However,
when the bias $b$ defined by (\ref{047}) is small, the dynamical process is
governed by fluctuations which become macroscopic. Even if the initial
magnetization is exactly defined $\left(  \delta_{0}=0\right)  $ the diffusion
term in (\ref{017}) produces a width (\ref{039}) which eventually leads to a
flat distribution, as illustrated by (\ref{052}). It is only in the biased
regime, which leads the magnetization to a single value $+m_{\mathrm{F}}$ with
nearly unit probability, that fluctuations play little role for large $N$, and
that the time-dependent mean-field equation (\ref{023}) is sufficient to
describe the dynamics. When the bifurcation is active, the intuitive idea that
the variable $m$, because it is macroscopic, should display relative
fluctuations small as $1/\sqrt{N}$ becomes wrong except near the initial time
and near the final equilibrium states.

The two regimes are characterized by the final situation: either non-vanishing
probabilities $\mathcal{P}_{+}$ and $\mathcal{P}_{-}$ for the two alternatives
$+m_{\mathrm{F}}$ and $-m_{\mathrm{F}}$, or a single choice only. Everything
takes place, like for Buridan's ass, as if the process were governed by final
causes: the behavior is deterministic if the target is unique; it displays
large uncertainties in the dynamics if hesitation may lead to one target or
the other. In fact, the singularity of the problem arises from the structure
of (\ref{017}), which contains a deterministic drift term involving the
velocity $v\left(  m\right)  $ and a diffusion term. Small as $1/N$, the
latter term does not contribute significantly when the bias is sufficient to
determine a single outcome ($\left\vert \lambda\right\vert \gg1$); it becomes
essential when the bifurcation is active ($\lambda$ finite or small), because
it governs the right-hand side of (\ref{017}) near the fixed points of the
drift motion, where $v\left(  m\right)  $ vanishes. The long duration of the
relaxation and the large uncertainties in its dynamics then reflect in a
probabilistic language the slowness of the pure drift motion around the
unstable fixed point, and the long and random delay needed to set $m$ into
motion. The very direction of the drift, which will be reflected in the sign
of the final magnetization, is thus quite sensitive to the perturbation caused
by the diffusion term, which altogether determines the long-time behavior.


\begin{thebibliography}{9}
\bibitem{12} The last author has not witnessed such an event.
\end{thebibliography}
\end{document}